\documentstyle[11pt,epsfig,citesort,subeqn]{article}

\def\vec#1{\mbox{\boldmath $#1$}}
\renewcommand{\theequation}{\thesection.\arabic{equation}}

\begin{document}

\begin{flushright}
TUM/T39-97-05 \\
SNUTP-97-023 \\
hep-ph/9703219
\end{flushright}

\begin{center}

{\Large\bf Solitons Bound to Heavy Mesons} \\

\vskip 2cm

Yongseok Oh\footnote{e-mail address : yoh@physik.tu-muenchen.de} \\

\vskip 2ex

{\it Institut f\"ur Theoretische Physik,
     Physik Department, Technische Universit\"at M\"unchen,
     D-85747 Garching, Germany} 

\vskip 2ex

Byung-Yoon Park\footnote{e-mail address : bypark@nsphys.chungnam.ac.kr} \\

\vskip 2ex

{\it Department of Physics, Chungnam National University,
     Daejeon 305-764, Korea}


\vskip 2cm {\bf ABSTRACT} \\   \begin{quotation} 

We improve the bound state approach of the Skyrme model applied to the
heavy baryons by adopting a static heavy meson picture where the
soliton moves around the fixed heavy meson.
This allows to take into account the center of mass corrections in a
more consistent way.
The bound state masses so obtained are comparable to the experimentally 
observed $\Lambda_c$ and $\Lambda_c^*$ masses.
A loosely bound state of a soliton with an antiflavored heavy meson is found,
which leaves a possibility of the nonstrange pentaquark baryon(s).

\vskip 1cm

\noindent\noindent{PACS number(s): 12.39.Dc, 12.39.Hg, 14.20.Lq}

\end{quotation}

\end{center}

\newpage

\section{Introduction}

In the Skyrme model, heavy baryons containing a single charm or bottom 
quark can be described by bound states of soliton and the heavy meson
of the corresponding heavy flavor
\cite{RRS,JMW,JM-GLM,JMW2,NRZ,Syracuse,MOPR,OP3,HQSS}.
It was first developed to describe strange hyperons \cite{CK}
and was extensively applied to heavy baryons by Rho, Riska and 
Scoccola \cite{RRS}, for which the SU(3) symmetric Lagrangian 
was strongly modified to incorporate large flavor-symmetry breaking. 
Although qualitatively successful, because of the lack of the heavy
quark symmetry in their Lagrangian, the resulting heavy baryon spectra
could not be consistent with the required symmetry. 
Later, Jenkins, Manohar and Wise \cite{JMW} have resolved this
problem by adopting a proper heavy-meson effective Lagrangian
\cite{Wise} that incorporates both the chiral symmetry and the heavy
quark symmetry explicitly.
Based on the model Lagrangian of Ref. \cite{Wise}, we have carefully
examined this picture to study the heavy baryon states with higher spin
\cite{OPM2} and also to study nonstrange pentaquark exotic baryons
\cite{OPM3}%
\footnote{For an extension of this idea to the NJL soliton model, see, for
example, Ref. \cite{GWZR}.}.

These works started out in the large $N_c$ (number of color) and
large $m_Q$ (heavy quark mass) limit where the soliton and the heavy
mesons are infinitely heavy so that they sit on the top of each other.
Such a simple picture is not good enough to describe
quantitatively the real heavy baryons, especially the excited states. 
In a series of papers \cite{OPM1,OP1-2}, we have tried to improve
the model by incorporating the finite mass effects of the heavy mesons. 
We adopted an effective Lagrangian \cite{YCCLLY} constructed for the
heavy mesons of finite mass and solved exactly the equations of motion
of the heavy mesons derived from it. The results are comparable to
the quark model calculations in the charm and/or bottom sector.

However, we have worked in the conventional soliton-fixed frame by taking
infinite $N_c$ limit but with finite $m_Q$.
It raises the conceptual problem of `the tail wagging the dog' \cite{RRS}.
Actually such a problem has been present even in the bound state approach
to the strange hyperons.
Since the mass of the kaon is about one half of the soliton mass, a large
center of mass correction is expected.
In case of the bound state approach to heavy baryons, the situation is
worse; the heavy meson is much heavier than the soliton --- $D$ ($B$)
meson masses are around 2 GeV (5 GeV) whereas the soliton mass is about
1 GeV.
Therefore it is obviously questionable that the former moves around the
latter.
A phenomenological consequence is that the heavy quark flavor symmetry
appears broken much more than expected \cite{OP1-2}.
As a crude way of estimating the center of mass correction, the reduced
mass of the heavy-meson--soliton system was used instead of the heavy
meson mass \cite{OP1-2,SS}, which results in a quite different spectrum.

In the real world, however, since the heavy mesons are much heavier than
the soliton, the soliton-fixed frame may not be a good starting point if
one wants to treat any center of mass corrections in a perturbative way.  
It would be more reasonable to start from the opposite picture where the
soliton moves in the heavy-meson-fixed frame, i.e., infinite $m_Q$ limit
but with finite $N_c$.
Work along this direction was first developed in Ref. \cite{JMW2}, where
the interaction between the heavy meson and the soliton was approximated
to that of the simple harmonic oscillator.
The effect of the kinetic motion to the ground state energy is included
through the zero point energy. 
In Ref. \cite{CW}, the same harmonic approximation is adopted to describe
the excited states $\Lambda^*_Q$. 
However, the crudeness of the approximate potential leads to the
$\Lambda^*_Q$ states lying high above the experimental observation.

In this paper, we improve the harmonic approximation of
Refs. \cite{JMW2,CW} by obtaining the equation of motion for the soliton
in a more realistic way. 
In the conventional soliton-fixed picture, the equations of motion for
the heavy mesons have been easily derived from the given heavy meson
effective Lagrangian after substituting the stationary hedgehog ansatz 
for the soliton configuration centered at the origin.   
On the contrary, in the picture of fixed heavy meson, we already have a 
known wavefunction of the heavy meson which should be localized at the
origin. 
So the static hedgehog ansatz is necessarily modified to generate the
kinetic motion of the soliton.
For this purpose, we introduce the collective coordinates representing
the translational motion of the soliton. 
The classical Lagrangian of the collective coordinates is then obtained
when the given effective Lagrangian density is integrated over the space
with the heavy meson fields localized at the origin and the chiral field
of the soliton configuration.
Canonical quantization, then, leads to a nonrelativistic Schr\"{o}dinger
equation for the motion of the soliton interacting with the heavy meson
sitting at the origin. 
This process is described in Sec. 2. 

In Sec. 3, we compare the resulting equation of motion with that
of the harmonic oscillator approximation. 
The Schr\"{o}dinger equation is solved numerically to obtain the 
bound states. 
The model yields a few bound states of soliton and heavy mesons
and the results are compared with the available experimental masses of
$\Lambda_c$ and $\Lambda_c^*$ states. 
We also investigate the bound state(s) of the soliton to the 
antiflavored heavy mesons, which corresponds to the pentaquark
($\bar{Q}q^4$) exotic baryons. 
Section 4 contains a summary and conclusion and we provide detailed
expressions in Appendices.

\setcounter{equation}{0}
\section{Soliton motion in the heavy meson fixed frame}

We start from the heavy meson effective Lagrangian of leading 
order in $1/m_Q$ expansion \cite{Wise}, which reads 
\begin{subequations}
\label{Leff}
\begin{equation}
{\cal L} = {\cal L}_M
 + i v_\mu \mbox{Tr} [ H (\partial^\mu + V^\mu) \bar{H}] 
 + g_Q \mbox{Tr}(H\gamma^\mu \gamma_5 A_\mu \bar{H}),
\end{equation}
where ${\cal L}_M$ is an $\mbox{SU(2)} \times \mbox{SU(2)}$ chiral
Lagrangian governing the dynamics of the Goldstone bosons.
For a simplicity, we will take the Skyrme model Lagrangian for ${\cal L}_M$; 
\begin{equation}
{\cal L}_M = 
\frac{f_\pi^2}{4} \mbox{Tr} (\partial_\mu U^\dagger \partial^\mu U)
+ \frac{1}{32e^2} \mbox{Tr} [ U^\dagger \partial_\mu U, 
U^\dagger \partial_\nu U]^2, 
\end{equation}
with the pion field represented by 
$U=\exp(i\vec{\tau}\cdot\vec{\pi}/f_\pi)$. 
The vector and axial-vector fields $V_\mu$ and $A_\mu$ are defined as 
\begin{eqnarray}
V_\mu &=& \frac{1}{2}(\xi^\dagger \partial_\mu \xi
 + \xi \partial_\mu \xi^\dagger), \\
A_\mu &=& \frac{i}{2}(\xi^\dagger \partial_\mu \xi
 - \xi \partial_\mu \xi^\dagger), 
\end{eqnarray}
in terms of $\xi = \exp(i\vec{\tau}\cdot\vec{\pi}/2f_\pi)$.
The heavy meson fields are combined into a $4 \times 4$ matrix field $H(x)$ as 
\begin{eqnarray}
H = \frac{1+v\hskip-0.5em\slash}{2} (P^*_\mu \gamma^\mu - P\gamma_5) \quad
\mbox{and} \quad
\bar{H} = \gamma_0 H^\dagger \gamma_0,
\end{eqnarray}
with the help of the standard Dirac matrices $\gamma_\mu$ and $\gamma_5$. 
Here, $P$ and $P^*_\mu$ are the heavy pseudoscalar and vector mesons 
moving with 4-velocity $v_\mu$, respectively.
This Lagrangian contains three physical constants, the pion decay constant
$f_\pi$, the Skyrme parameter $e$ and the coupling constant $g_Q$ of heavy
mesons.
We will take them as free parameters varying around their empirical values
($f_\pi \sim 93$ MeV and $e \sim 4.8$) together with a nonrelativistic 
quark model prediction ($g_Q = -\frac34$). 
One can easily check the invariance of the Lagrangian (\ref{Leff}) under
the $\mbox{SU(2)}_L \times \mbox{SU(2)}_R$ chiral transformation 
\begin{eqnarray}
\begin{array}{l}
U \rightarrow L U R^\dagger \quad
(\xi \rightarrow L \xi h^\dagger = h \xi R^\dagger), \\
H \rightarrow H h^\dagger. 
\end{array} 
\end{eqnarray}
\end{subequations}

In the conventional bound state approaches, the static soliton
configuration centered at the origin 
$U_0(\vec{r}) = \exp [i \vec{\tau}\cdot\hat{\vec{r}} F(r)]$ is substituted 
into the Lagrangian and the resulting equations of motion for the
heavy-flavored mesons (or kaons) moving in the static potentials arising
from the soliton are solved to find the bound states. 
This `soliton--fixed' picture supported from the large $N_c$ arguments 
could be a good starting point for the strange baryons, which could be
improved by proper center of mass corrections later. 
In case of the bound state approach to heavy baryons, however, the
situation is quite different.
The observed heavy mesons $D$ and $D^*$ or $B$ and $B^*$ are definitely
much heavier than the soliton. 
In this case, more reasonable starting point would be the 
heavy-meson--fixed picture, where the soliton moves around the 
heavy meson fixed at the origin and the corrections from the finite heavy
meson mass could be included perturbatively.

In the heavy-meson--fixed frame, the spatial part of the 
heavy meson wavefunction is already determined by a well-localized 
function at origin such as $\delta^3(\vec{r})$.
To generate the motion of the soliton, we substitute an ansatz \cite{BTW}
\begin{equation}
U(\vec{r},t) = U_0(\vec{r}-\vec{x}(t)),
\end{equation}
into the Lagrangian for the soliton configuration with its center at
$\vec{x}(t)$.    
Carrying out the integration over $\vec{r}$, we obtain the classical 
Lagrangian for the collective variables $\vec{x}(t)$ as%
\footnote{We treat the soliton motion nonrelativistically.}
\begin{subequations}
\label{Lx}
\begin{equation}
L = \int d^3 r {\cal L} = \textstyle M_s + \frac12 M_s \vec{u}^2 - 
\displaystyle
 \frac{2\upsilon(x)}{x}\vec{u}\cdot(\vec{I}_h \times \vec{x}) - V_I, 
\end{equation}
where 
\begin{equation}
V_I = 2g_Q [ a_1(x) \vec{S}_{\ell h} \cdot \vec{I}_h 
+ a_2(x) \vec{S}_{\ell h} \cdot \hat{\vec{x}} \,
\vec{I}_h \cdot \hat{\vec{x}} ],
\end{equation}
and $\vec{u}=d\vec{x}/dt$ is the velocity of the soliton and 
$M_s$ the soliton mass. (From now on, we will drop the trivial 
soliton mass term from the Lagrangian.) 
$\vec{S}_{\ell h}$ and $\vec{I}_h$ are the light quark spin and
isospin operator of the heavy mesons, respectively.
The radial functionals $\upsilon(x)$, $a_1(x)$ and $a_2(x)$
are defined as  
\begin{equation}
\upsilon(x) = \frac{\sin^2(F/2)}{x}, \quad
a_1(x) = \frac{\sin F}{x} \quad \mbox{and} \quad
a_2(x) = \frac{dF}{dx} - \frac{\sin F}{x}. 
\label{Lxb}
\end{equation}
\end{subequations}

The classical Hamiltonian is obtained by taking the Legendre 
transformation of the Lagrangian (\ref{Lx}),
\begin{equation}
H_{\rm cl} = \frac{1}{2M_s} \left\{ \vec{p}^2 
+ \frac{4\upsilon}{x} \vec{I}_h \cdot \vec{L} 
+ 2 \upsilon^2 \right\} + V_I(x), 
\end{equation}
with the conjugate momentum $\vec{p}$ defined as 
\begin{equation}
\vec{p} = \frac{\partial L}{\partial \vec{u}} 
= M_s \vec{u} - \frac{2\upsilon}{x} (\vec{I}_h \times \vec{x}).
\end{equation}
Quantization of this collective motion leads us to the 
Schr\"{o}dinger equation in the form of \cite{JMW2}
\begin{equation}
\left\{ -\frac{1}{2M_s} \vec{\nabla}^2 + \frac{1}{2M_s} \left[ 
\frac{4\upsilon}{x} \vec{I}_h\cdot \vec{L} + 2\upsilon^2 \right]
+ V_I \right\} \Psi_n(\vec{x}) = \varepsilon_n \Psi_n(\vec{x}).
\label{SchEq}\end{equation}
The wavefunction $\Psi(\vec{x})$ depends not only on the collective
coordinates $\vec{x}$ but also on the light-quark spin $s_{\ell h}$
and the isospin $i_h$ of the heavy mesons.

Since the Hamiltonian commutes only with the `light-quark grand spin'
operator defined by $\vec{K}_\ell = \vec{L} + \vec{S}_{\ell h} + \vec{I}_h$, 
the eigenstates can be classified by the corresponding quantum numbers 
($k_\ell, m$) and the wavefunction can be written as 
\begin{equation}
\Psi(\vec{x}) = \sum_{i} \psi_i(x) \,
\langle \hat{\vec{x}}|k_\ell, m \rangle_{i}, 
\end{equation} 
where the summation $i$ runs over the possible ways of combining  
the grand spin eigenstates $\langle \hat{\vec{x}}|k_\ell, m\rangle$ from  
orbital angular momentum basis $Y_{\ell, m_\ell}(\hat{\vec{x}})$, 
light-quark spin basis $|s_{\ell h}=\frac12, \pm\frac12\rangle$ 
and the isospin basis $|i_h=\frac12, \pm\frac12\rangle$ of the 
heavy mesons. 

For a given $k_\ell(\neq 0)$, there are four possible eigenstates 
(see Appendix A) and, because of their different parity, only two 
states are mixed for a given parity. 
The wavefunction, then, can be written explicitly as 
\begin{equation}
\Psi(\vec{x}) =
   \psi_1(x) \, \langle \hat{\vec{x}}|k_\ell, m\rangle_1
 + \psi_2(x) \, \langle \hat{\vec{x}}|k_\ell, m\rangle_2, 
\end{equation}
for the states of parity $\pi = -(-1)^{k_\ell}$, and 
\begin{equation}
\Psi(\vec{x}) =
   \psi_3(x) \, \langle \hat{\vec{x}}|k_\ell, m\rangle_3
 + \psi_4(x) \, \langle \hat{\vec{x}}|k_\ell, m\rangle_4, 
\end{equation}
for the states of parity $\pi = +(-1)^{k_\ell}$ including the intrinsic
parity of heavy mesons.

Then, the Schr\"{o}dinger equation (\ref{SchEq}) is reduced to 
coupled ordinary differential equations for the radial functions 
$\psi_i(x)$ as 
\begin{subequations}
\label{xeq}
\begin{equation}
\left\{ -\frac{1}{2M_s} \left[ \frac{d^2}{dx^2}
 + \frac{2}{x} \frac{d}{dx} \right] \vec{1}_2
 + V_{\rm eff} \right\} \psi_n(x)
 = \varepsilon_n \psi_n(x), 
\end{equation}
with the $2 \times 2$ unit matrix $\vec{1}_2$, where
$V_{\rm eff}$ is in the form of $2 \times 2$ matrix 
\begin{equation}
V_{\mbox{\scriptsize eff}}(x)  = 
\left( \begin{array}{cc} 
V_{11} & V_{12} \\ V_{21} & V_{22} \end{array} \right) 
\mbox{ or } 
\left( \begin{array}{cc} 
V_{33} & V_{34} \\ V_{43} & V_{44} \end{array} \right),  
\end{equation}
and the corresponding radial function $\psi(x)$ is
\begin{equation}
\psi(x) = 
\left( \begin{array}{c} \psi_1(x) \\ \psi_2(x) \end{array} \right) 
\mbox{ or }
\left( \begin{array}{c} \psi_3(x) \\ \psi_4(x) \end{array} \right). 
\end{equation}
\end{subequations}
The explicit expressions for the potentials $V_{ij}$ can be found
in Appendix A.

\setcounter{equation}{0}
\section{Bound states of soliton}

Before solving the Schr\"{o}dinger equation numerically, let us 
compare Eq. (\ref{xeq}) with the harmonic approximation of 
Refs. \cite{JMW2,CW}.
By making use of the fact that the soliton function $F(x)$ behaves 
near its center as 
\begin{equation}
F(x) = \pi + F^\prime(0) x 
 + \textstyle\frac{1}{6} F^{\prime\prime\prime}(0) x^3 + O(x^4), 
\end{equation}
we can expand the radial functionals of Eq. (\ref{Lxb}) as
\begin{equation}
\begin{array}{l}
\displaystyle
\upsilon(x) = \frac{1}{x}
 - \textstyle \frac14[F^\prime(0)]^2 x + O(x^3), \\[3mm]
a_1(x) = - F^\prime(0) - \frac16 \{ F^{\prime\prime\prime}(0) 
- [F^\prime(0)]^3 \} x^2 + O(x^3), \\[3mm]
a_2(x) = 2F^\prime(0) + \frac16 \{4F^{\prime\prime\prime}(0)
 - [F^\prime(0)]^3 \} x^2 + O(x^3). 
\end{array} 
\label{RF}
\end{equation}

In the extreme limit of $N_c \rightarrow \infty$, the soliton 
is also infinitely heavy so that it sits on the top of the heavy mesons.
The binding energy of the system is the value of the potentials 
at the origin. 
Neglecting the centrifugal energy coming from the kinetic motion 
of the soliton, we obtain
\begin{equation}
\begin{array}{c}
\displaystyle
\tilde{V}_{\mbox{\scriptsize eff}}(0)
 = \textstyle\frac12 g_Q F^\prime(0) \displaystyle  
\left( \begin{array}{cc} 1 & 0 \\ 0 & 1 \end{array} \right),
\\[3mm]
\displaystyle
\tilde{V}_{\mbox{\scriptsize eff}}(0)
= - \frac{g_Q F^\prime(0)}{2(2k_\ell+1)}
\left( \begin{array}{cc} 2k_\ell+3 & -4\sqrt{k_\ell(k_\ell+1)} \\
-4\sqrt{k_\ell(k_\ell+1)} & 2k_\ell-1 \end{array} \right),
\end{array}
\end{equation}
for $\pi=-(-1)^{k_\ell}$ states and $\pi = +(-1)^{k_\ell}$ states, 
respectively.
(In order to distinguish the potentials obtained with no centrifugal
energy, we use the tilde on $V$.)
Diagonalization of those potential matrices yields the results
of Refs. \cite{OPM2,OPM3}; that is,
degenerate heavy-meson--soliton bound states 
with binding energy $-\frac32 g_Q F^\prime(0)$ 
and antiflavored-heavy-meson--soliton bound states 
with binding energy $-\frac12 g_Q F^\prime(0)$. 
These degeneracies are of course the artifacts of ignoring the 
kinetic motion, especially, the centrifugal energy.   

In case of $k_\ell^\pi = 0^+$, we have only one state 
$\langle \hat{\vec{x}} |k_\ell=0, 0\rangle_3$. 
Although this state is made of $\ell=1$ angular momentum basis, the
associated singularity in the centrifugal energy is canceled by the
same one of $\upsilon(x)$ in the potential, and the resulting
effective potential is finite at the origin. 
Consequently, it provides the lowest energy bound state, if any.
If we take the harmonic approximation of Eq. (\ref{RF}), the potential
energy can be written as
\begin{subequations}
\begin{equation}
V_{33}(x) \approx V_0 + \textstyle\frac12 \kappa x^2,
\label{hoa}\end{equation}
where 
\begin{eqnarray}
V_0 &=& -\textstyle\frac32 g_Q F^\prime(0), \\
\textstyle 
\kappa &=& g_Q \{\frac13 [F^\prime(0)]^3
 - \frac56F^{\prime\prime\prime}(0) \}.
\end{eqnarray}
\end{subequations}
In Table \ref{t1}, we list the values of $V_0$ and $\kappa$ 
obtained by using Eq. (\ref{hoa}) with conventional 
parameters of the Skyrme model.

\begin{table} 
\caption{\sl $V_0$ and $\kappa$ evaluated with 
conventional Skyrme model parameters. Energies are given in MeV 
unit and $\kappa$ in MeV$^3$ unit.}
\begin{center} 
\begin{tabular}{ccccrrrrr}
\hline\hline
Set & $f_\pi$ & $e$ & $g_Q$ & \multicolumn{1}{c}{$M_s$} 
& \multicolumn{1}{c}{$V_0$} & 
\multicolumn{1}{c}{$\kappa$} & \multicolumn{1}{c}{$E_b$} & 
\multicolumn{1}{c}{$E_e$} \\
\hline
  (I) & 64.5 & 5.45 & $-0.75$ & 867 & $ -791$ & $(575)^3$ & $ -88$ & 468 \\
 (II) & 64.5 & 5.45 & $-1.00$ & 867 & $-1055$ & $(632)^3$ & $-246$ & 540 \\
(III) & 93.0 & 4.82 & $-0.75$ & 1105 & $-1009$ & $(733)^3$ & $-113$ & 597 \\
 (IV) & 93.0 & 4.82 & $-1.00$ & 1105 & $-1345$ & $(807)^3$ & $-311$ & 690 \\
\hline
 Expt. & 93.0 & $-$  &   $-$   & 867 & $-1095$ & $(437)^3$ & $-630$ & 310 \\
\hline \hline
\end{tabular}
\end{center}
\label{t1}\end{table} 

In this harmonic approximation, the eigenstates can be easily obtained as 
\begin{equation}
\varepsilon_n = V_0
+ \sqrt{\frac{\kappa}{M_s}} \left( n + \textstyle \frac32 \right),
\label{enho}\end{equation}
where $n$ is an even integer number. 
The lowest energy state of $n$=0 is the ground state which has a 
negative energy, which corresponds to the binding energy of the 
$\Lambda_Q$ baryon. From Eq. (\ref{enho}), we can write 
the mass formula for $\Lambda_Q (\frac12^+)$ baryons as  
\begin{equation}
m_{\Lambda_Q} = m_N + m_H
 + V_0 +  \sqrt{\frac{\kappa}{M_s}} \left( N + \textstyle \frac32 \right) , 
\end{equation}
where we have included the mass of the heavy mesons $m_H$ and the 
nucleon mass $m_N$ (soliton mass plus the energy from collective rotation).  
The kinetic motion of the soliton with finite mass is included through
the zero point energy of $\frac32\sqrt{\kappa/M_s}$.  
The artificial degeneracy in the model of $N_c, m_Q \rightarrow \infty$
is thus removed, while only the ground state is bound.

In Ref. \cite{CW}, this harmonic approximation of the potential was 
straightforwardly extended to $k_\ell > 0$ states by simply adding the 
centrifugal energy. 
The tensor couplings of $V_{12}$ ($V_{34}$) between the 
$|k_\ell, m\rangle_1$ and $|k_\ell, m\rangle_2$ states
($|k_\ell, m\rangle_3$ and $|k_\ell, m\rangle_4$ states)
was not taken into account in this approximation. 
Then, the potential gives rise to an infinite tower of eigenenergies 
of Eq. (\ref{enho}) for any integer $n$.
The first excited $k_\ell^\pi=1^-$ states which is $\sqrt{\kappa/M_s}$ 
above the ground state could be interpreted as $\Lambda_Q^*(\frac12^-)$ and 
$\Lambda_Q^*(\frac32^-)$ states. 
In order to fit the observed masses of 
$\Lambda_c$ (2285 MeV), $\Lambda_c^*$($\frac12^-$) (2593 MeV) and 
$\Lambda_c^*$($\frac32^-$) (2625 MeV) \cite{PDG96,Lam:exct},   
we need to have the binding energy $E_b$ 
and the excitation energy $E_e$ as 
\begin{equation} 
\begin{array}{l}
E_b = V_0 + \frac{3}{2} \sqrt{\kappa/M_s} = -630 \mbox{ MeV}, \\[3mm]
E_e = \sqrt{\kappa/M_s} = 310 \mbox{ MeV}, 
\end{array}
\end{equation}
which are regarded as the `experimental' values in Table \ref{t1}.
As can be seen in Table \ref{t1}, under the harmonic 
approximation, it is not easy to fit the experimental values 
by adjusting the parameters of the Skyrme model 
around their empirical values in a reasonable range, which was also
pointed out in Ref. \cite{SS}.

\begin{figure}[t]
\centerline{\epsfig{file=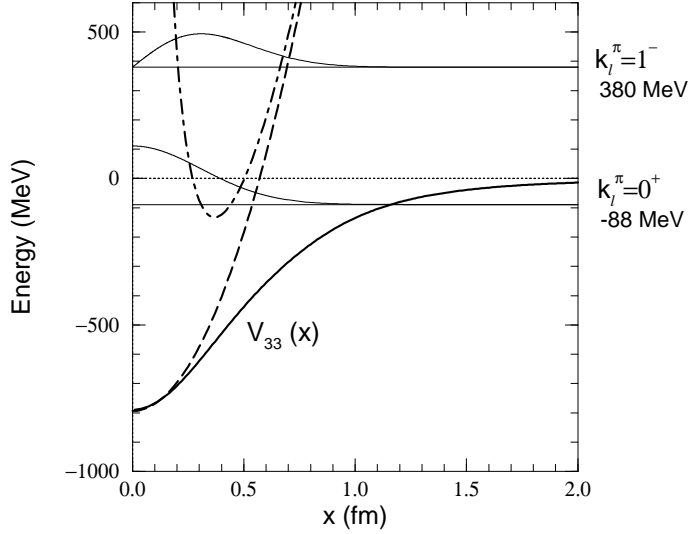, width=10cm}}
\caption{\sl Potentials and wavefunctions in the harmonic oscillator
approximation with the parameter set (I).
The solid line is the exact form of $V_{33}$ and the
long dashed line is its harmonic approximation. 
The potential of $k^\pi_\ell=1^-$ state including the centrifugal 
energy is given by the dot-dashed line. The corresponding
wavefunctions are also presented.}
\end{figure}

Shown in Fig. 1 is the harmonic approximation (long dashed line) 
of the potential $V_{33}(x)$ in comparison with the exact one (solid line).  
The dot-dashed curve is the effective potential in the harmonic approximation 
with the centrifugal energy 
$\ell_{\rm eff}(\ell_{\rm eff}+1)/(2M_s x^2)$
with $(\ell_{\rm eff}=1)$ included. 
The lowest energy levels of the corresponding harmonic potentials 
are explicitly presented with their wavefunctions in the parameter
set (I).
First of all, compared with the exact one, the approximate potential is
too stiff. 
Because of this crudeness of the harmonic oscillator potential, 
the zero point fluctuation and the excitation energy become too large.
This is crucial for the excited states whose wavefunctions are spread
over a wide range.
Consequently, the binding energy of the $\Lambda_Q$ appears too small  
and the first excited $\Lambda^*_Q$ state locates too high above the
experimental observations for all the parameter sets adopted in
Table~\ref{t1}. 

\begin{figure}[t]
\centerline{\epsfig{file=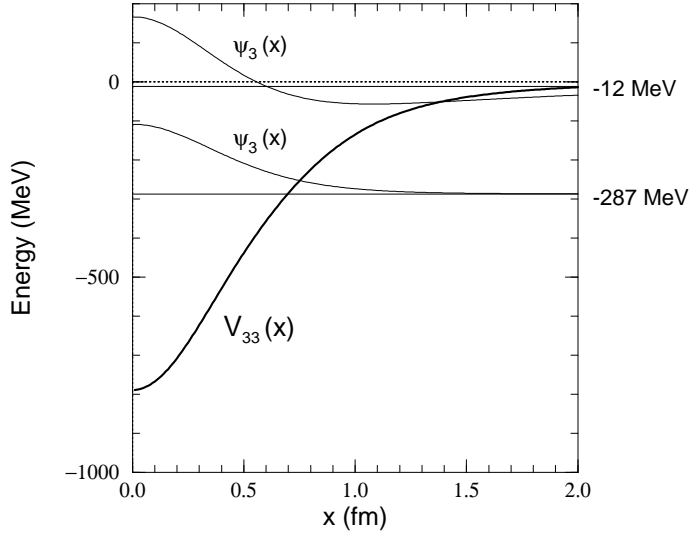, width=10cm}}
\caption{\sl Solutions for the $k_\ell^\pi=0^+$ bound states with the
parameter set (I).
Thick solid line is the potential $V_{33}$ and
the solid lines are the wavefunctions of the ground and radially
excited states.}
\end{figure}

Now, let us investigate the exact solutions. 
In Fig. 2, we present two $k_\ell^\pi=0^+$ bound states and their
wavefunctions obtained with the parameter set (I). 
Comparing with those of the harmonic approximation (Fig. 1), 
the exact wavefunctions are more expanded so that the eigenstates
get more binding. 
In Fig. 3, solid lines are the potential energies $V_{ij}(i,j=3,4)$ 
for the $k_\ell^\pi=1^-$ state. 
At a first glance, there does not appear any attractive 
part in each potential, with the exception of $V_{44}$ which is however
not strong enough to support a bound state. 
To understand the binding mechanism, we diagonalize 
the potential matrix at each position $x$:
\begin{equation}
S(x) V_{\mbox{\scriptsize eff}} S^\dagger(x)
= \mbox{diag} (V^+, V^-),
\label{st}\end{equation}
where 
\begin{equation}
V^\pm(x) = \textstyle \frac12 [ (V_{33}+V_{44}) 
\pm \sqrt{(V_{33}-V_{44})^2 + 4V_{34}^2} ].
\end{equation}
These $V^\pm$ are presented by dashed lines in Fig. 3. 
The off-diagonal potential $V_{34}$ pushes down $V_{44}$ so that it 
becomes sufficiently attractive to support a bound state.
Of course, one should {\em not} use one of the diagonalized 
potentials, $V^+$ or $V^-$, to obtain the eigenstates of the 
equation of motion because the similarity transformation matrix $S(x)$ 
of Eq. (\ref{st}) is local. Instead, one should follow the general 
procedure to solve the coupled equations as described in Appendix B. 
The resulting bound state is shown in the inner box of Fig. 3 
with its wave functions $\psi_3(x)$ and $\psi_4(x)$
(solid lines). There, $V^-$ is drawn only as a guiding line.

\begin{figure}[t]
\centerline{\epsfig{file=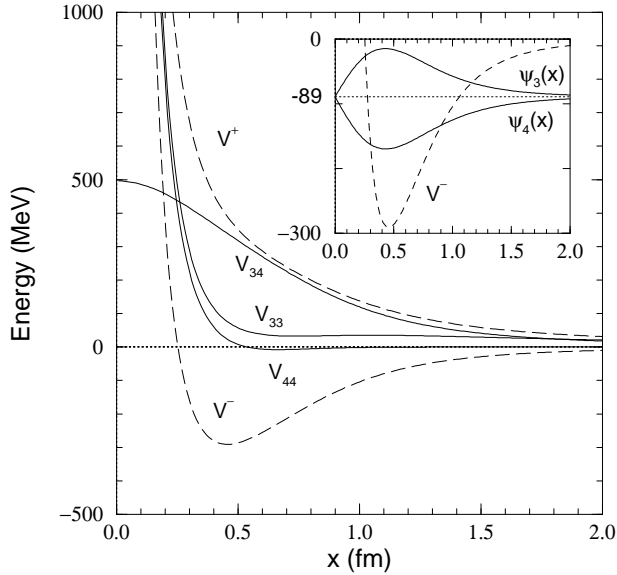, width=10cm}}
\caption{\sl Solutions for the $k_\ell^\pi=1^-$ bound states
with parameter set (I).
The solid lines are potentials $V_{ij}$ ($i,j=3,4$) and the dashed lines
are $V^\pm$.
The wavefunctions are given in the inner box with $V^-$.}
\end{figure}

We summarize our numerical results in Table \ref{t3}.
We list the bound states of the model with four 
different parameter sets (I)--(IV). 
In each case, we could find two bound states in the 
$k_\ell^\pi = 0^+$ channel and one in the $1^-$ channel.
This should be contrasted with the results of the soliton-fixed frame of
Ref. \cite{OP1-2}, where many bound states appear as we increase
the heavy meson mass.
The coupling constant predicted by the nonrelativistic quark model
($|g_Q|=\frac34$) yields too small binding energies compared with the 
experimental values.
By increasing its magnitude to 1, which may represent the role of light
vector mesons \cite{SS}, we could obtain reasonable values. 
Comparing with the parameter set (I) and (II) with $f_\pi=64.5$ MeV and 
$e=5.45$, set (III) and (IV) of $f_\pi=93$ MeV and $e=4.82$
result in more strongly bound states. 
It is mainly due to the different shapes of the potentials and partly
due to the heavier soliton mass ($M_s=1105$ MeV) of the latter 
which slightly exceeds the required mass 867 MeV to fit the nucleon 
mass. In adopting these parameter sets, we account for the 
Casimir energy \cite{MK} that will reduce down the final soliton mass.   

\begin{table} 
\caption{\sl Numerical results on the bound states. 
Energies are given in MeV unit.}
\vskip 3mm \begin{center}
\begin{tabular}{l|rrrr|r}
\hline\hline
($n$, $k_\ell^\pi$) & \multicolumn{1}{c}{Set I} & 
\multicolumn{1}{c}{Set II} & \multicolumn{1}{c}{Set III} 
&  \multicolumn{1}{c|}{Set IV} &  \multicolumn{1}{c}{Exp.} \\
\hline
(0, $0^+$) & $-287$ & $-461$ & $-366$ & $-588$ & $-610$ \\
(1, $0^+$) &  $-12$ &  $-62$ &  $-15$ &  $-79$ &  \multicolumn{1}{c}{$-$} \\
(0, $1^-$) &  $-89$ & $-196$ & $-113$ & $-250$ & $-320$ \\
\hline
(0, $1^+$)$^*$ & $-17$ & $-54$ & $-21$ & $-69$  &  
\multicolumn{1}{c}{$-$} \\
\hline\hline
\multicolumn{6}{c}{$^*$ bound state of soliton to antiflavored heavy meson} 
\end{tabular}
\end{center}
\label{t3}\end{table}

\begin{figure}[t]
\centerline{\epsfig{file=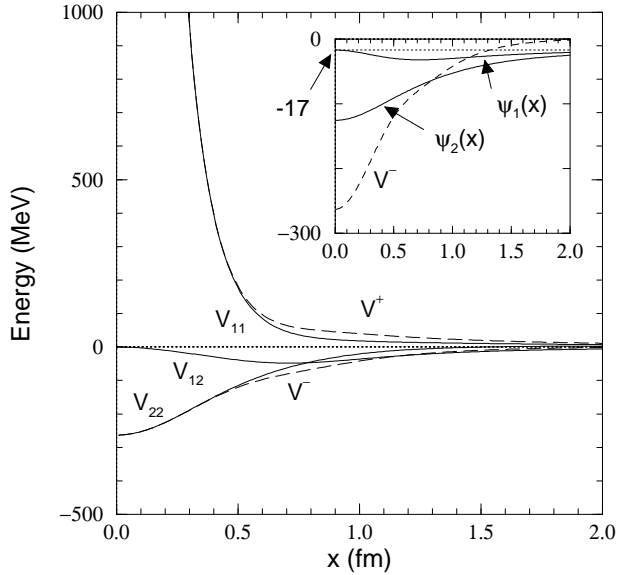, width=10cm}}
\caption{\sl Solutions for the $k_\ell^\pi=1^+$
soliton--antiflavored-heavy-meson bound states with the parameter set (I).
The solid lines are potentials $V_{ij}$ ($i,j=1,2$) and the dashed lines
are $V^\pm$.
The wavefunctions are given in the inner box with $V^-$.}
\end{figure}

The last line of Table \ref{t3} is the result for soliton bound state
to the antiflavored heavy meson,   which corresponds to the pentaquark
($\bar{Q}q^4$) exotic baryon in the quark model where $q=u,d$ quarks.
It can be searched simply by changing the sign of the coupling constant
$g_Q$ \cite{OPM3,RS}.  
We could find only one bound state in the $k_\ell^\pi = 1^+$ channel. 
Shown in Fig. 4 are the potentials $V_{ij}(i,j=1,2)$ for
the soliton--antiflavored-heavy-meson interactions (solid lines), 
the diagonalized one as in Fig. 3 (dashed lines) and
the wavefunctions $\psi_1(x)$ and $\psi_2(x)$ of the bound state
(inner box). 
In case of the nonrelativistic quark model value of $g_Q$, 
the state is bound too weakly to survive after the $1/m_Q$ 
corrections. 
However, with a coupling constant $|g_Q|\sim 1$ which yields reasonable
soliton--heavy-meson bound states, the binding energy of the bound state
of soliton and antiflavored-heavy-meson is somewhat large and
it still leaves a possibility of loosely bound {\em nonstrange}
pentaquark baryons.

\section{Summary and Conclusion}

In this paper, we have carried out the bound state approach of the 
Skyrme model to heavy baryons in a more realistic way.
We obtain the bound states of the soliton and heavy meson 
in the fixed heavy-meson picture. 
It improves the harmonic approximation of Refs. \cite{JMW2,CW} 
by deriving the soliton--heavy-meson interacting potential 
in a more consistent way from a given effective Lagrangian. 
First we introduce the collective coordinates for the 
motion of the soliton. Then, the classical Lagrangian for these
coordinates is obtained by carrying out the integral over the space
with localized heavy meson fields. The nonrelativistic 
Scr\"{o}dinger equation is obtained by quantizing the motion.
The resulting bound states have larger binding energies than 
in the harmonic oscillator approximation. 
Apart from the ground state with $k_\ell^\pi=0^+$, 
we could find the first excited state with negative parity 
in the $k_\ell^\pi=1^-$ channel. 
With the value of $g_Q \sim 1$, which is slightly larger than the 
nonrelativistic quark model prediction, those resulting 
bound states are comparable to the experimentally observed 
$\Lambda_c(\frac12^+)$ and $\Lambda_c^*(\frac12^-, \frac32^-)$
states.

This work also improves the work of Refs. \cite{OP1-2} where the bound
state approach was carried out in the conventional soliton-fixed 
picture. In this approach the heavy quark flavor symmetry is
rather strongly broken as expected. 
Since the $1/m_Q$ order terms are not included in the Lagrangian, 
the obtained bound states do not depend on the flavor 
of the heavy mesons at that level of approximation. 
If one turns on the $1/m_Q$ corrections, this degeneracy 
will be removed as much as the heavy flavor is broken by the 
$1/m_Q$ order terms in the Lagrangian.

We have obtained just those bound states which are eigenstates 
of the light-quark grand spin operator. 
They could be naively compared with the $\Lambda_Q(\frac12^+)$ and 
$\Lambda_Q^*(\frac12^-,\frac32^-)$ baryons. 
However, in order to provide those bound states with the correct quantum
numbers such as spin and isospin, we have to quantize the zero modes 
of the bound states under the SU(2) rotation in isospin space.
This procedure will enable us to check the model by comparing the results 
with the other observed heavy baryon states such as $\Sigma_Q$ and
$\Sigma^*_Q$ \cite{PDG96,DELPHI}. 
Work in this direction is in progress and will be reported elsewhere
\cite{OPM4}.


\section*{Acknowledgements}

We are grateful to D.-P. Min and M. Rho for useful discussions and to
W. Weise for careful reading of the manuscript with useful comments.
One of us (Y.O.) thanks the Alexander von Humboldt Foundation
for financial support.
This work was supported in part by the Korea Science and Engineering
Foundation through the SRC program.

\appendix

\setcounter{equation}{0}
\section{Matrix elements in $k_\ell$ basis}

In this Appendix, we give the explicit expressions for the potential
$V_{ij}$ of Eq. (\ref{xeq}).

The light-quark grand spin operator is defined as 
\begin{equation}
\vec{K}_\ell = \vec{L} + \vec{I}_h + \vec{S}_{\ell h},
\end{equation}
where $\vec{I}_h$ and $\vec{S}_{\ell h}$ are the isospin 
and light-quark spin operators of the heavy mesons 
and $\vec{L}$ is the orbital angular momentum operator of the soliton. 
It should be distinguished from the grand spin operator $\vec{K}$ 
that is the sum of the orbital angular momentum and all
kinds of the spin or isospin of the particles in the model. 
Note that in the definition of the operator $\vec{K}_\ell$
only the angular momentum operators 
of the light degrees of freedom are included.

The eigenstates of the light-quark grand spin operator can be obtained by 
linear combinations of direct product of the eigenstates of each angular
momentum operator in its definition:
\begin{equation}
\langle \hat{\vec{x}} | k_\ell, m\rangle_i = 
\sum_{m\mbox{'s}} \textstyle 
(\ell_i\, m_\ell\, \frac12\, m_t\,|\,\lambda_i\, m_\lambda)
(\lambda_i\, m_\lambda\, \frac12\, m_s\,|\, k_\ell\, m) 
Y_{\ell_i m_\ell}(\hat{\vec{x}})\,
|\frac12\, m_t\rangle_t\, |\frac12\, m_s\rangle,
\end{equation} 
with the help of the Clebsch-Gordan coefficients \cite{Edmonds}.
Here, we first combine $\vec{L}$ and $\vec{I}_h$ 
($\lambda$ and $m_\lambda$ are the corresponding quantum numbers 
of the eigenstates of $\vec{\Lambda} \equiv \vec{L}+\vec{I}_h$), 
and then combine the light-quark spin $\vec{S}_{\ell h}$. 
For a given $k_\ell(>0)$, there come out four eigenstates as 
listed in Table \ref{t4}. In case of $k_\ell=0$, there can be 
only two states $\langle \hat{\vec{x}} | k_\ell, m \rangle_{1,3}$.
In evaluating the total parity of the states as in Table \ref{t4},  
the negative intrinsic parity of the heavy mesons is taken into 
account.

\begin{table} 
\caption{\sl Four $\langle \hat{\vec{x}}|k_\ell m\rangle$ eigenstates for
a given $k_\ell$.}
\vskip 3mm \begin{center}
\begin{tabular}{c|c|c||c|c|c}
\hline\hline
\multicolumn{3}{c||}{parity $\pi=-(-1)^{k_\ell}$} &
\multicolumn{3}{c}{parity $\pi=+(-1)^{k_\ell}$} \\
\hline 
state & $\ell_i$ & $\lambda_i$ & state & $\ell_i$ & $\lambda_i$ \\
\hline
$\langle \hat{\vec{x}} | k_\ell, m \rangle_1$ & $\lambda_i - \frac12$ & 
  $k_\ell + \frac12$ & 
$\langle \hat{\vec{x}} | k_\ell, m \rangle_3$ & $\lambda_i + \frac12$ & 
  $k_\ell + \frac12$ \\
$\langle \hat{\vec{x}} | k_\ell, m \rangle_2$ & $\lambda_i + \frac12$ & 
  $k_\ell - \frac12$ & 
$\langle \hat{\vec{x}} | k_\ell, m \rangle_4$ & $\lambda_i - \frac12$ & 
  $k_\ell - \frac12$ \\
\hline \hline
\end{tabular} \end{center}
\label{t4}\end{table}

To obtain the potentials $V_{ij}(i,j$=1,2 or 3,4) of Eq. (\ref{xeq}),
we need to evaluate the expectation values of the operators,
$\vec{I}_h \cdot \vec{L}$, $\vec{I}_h \cdot\vec{S}_{\ell h}$ and 
$(\vec{I}_h \cdot \hat{\vec{x}}) (\vec{S}_{\ell h} \cdot \hat{\vec{x}})$, 
with respect to the eigenstates $| k_\ell, m\rangle_i(i=$1,2 or 3,4).
The first one is simple to evaluate the expectation values, 
since it can be rewritten as 
\begin{equation}
\vec{I}_h \cdot \vec{L} = \textstyle \frac12 
(\vec{\Lambda}^2 - \vec{L}^2 - \vec{I}^2_h) 
= \displaystyle \left\{
\begin{array}{ll} 
\ell/2, & \mbox{if $\lambda=\ell+\frac12$,} \\
-(\ell+1)/2, & \mbox{if $\lambda=\ell-\frac12$.} 
\end{array}
\right.
\end{equation}
Thus, it is trivial to obtain 
${}_i\langle \vec{I}_h \cdot \vec{L} \rangle_j$ as
\begin{equation}
\mbox{
\begin{tabular}{c|c|c}
$i,j$ & 1 & 2 \\[2mm]
\hline
1 & $\frac12 k_\ell$ & 0 \\[2mm]
\hline
2 & 0 & $-\frac12(k_\ell+1)$ \\
\end{tabular}
} 
\hskip 5mm
\mbox{
\begin{tabular}{c|c|c}
$i,j$ & 3 & 4 \\[2mm]
\hline
3 & $-\frac12 (k_\ell + 2)$ & 0 \\[2mm]
\hline
4 & 0 & $\frac12(k_\ell - 1)$ \\
\end{tabular}
} 
\end{equation}

The expectation values of the operator $\vec{I}_h \cdot \vec{S}_{\ell h}$
can have a simpler form if we obtained the eigenstates of the operator
$\vec{K}_\ell$ in another way; that is, by combining $\vec{S}_{\ell h}$ 
and $\vec{I}_h$ first 
($\vec{\Sigma}_\ell \equiv \vec{I}_h + \vec{S}_{\ell h}$ 
with the quantum numbers $\sigma$ and $m_\sigma$),  and then with $\vec{L}$.
Let $|\ell, \sigma; k_\ell, m)$ be such a state. 
Then, the eigenstates $|k_\ell, m \rangle_i$ can be related to the
eigenstates $|\ell, \sigma; k_\ell, m)$ by the Racah coefficients as 
\begin{equation}
\begin{array}{l} 
\displaystyle 
|k_\ell, m \rangle_1 = 
 \sqrt{ \frac{k_\ell+1}{2k_\ell+1} } |\ell=k_\ell, \sigma=0; k_\ell, m)
+\sqrt{ \frac{k_\ell}{2k_\ell+1} } |\ell=k_\ell, \sigma=1; k_\ell, m), \\
\displaystyle 
|k_\ell, m \rangle_2 = 
- \sqrt{ \frac{k_\ell}{2k_\ell+1} } |\ell=k_\ell, \sigma=0; k_\ell, m)
+\sqrt{ \frac{k_\ell+1}{2k_\ell+1} } |\ell=k_\ell, \sigma=1; k_\ell, m), \\
|k_\ell, m \rangle_3 = |\ell=k_\ell+1, \sigma=1; k_\ell, m), \\
|k_\ell, m \rangle_4 = |\ell=k_\ell-1, \sigma=1; k_\ell, m).
\end{array}
\end{equation}
By using the identity 
\begin{equation}
\vec{I}_h \cdot \vec{S}_{\ell h} = \frac12(\vec{\Sigma}_\ell^2 - \frac32),
\end{equation} 
we can obtain ${}_i\langle \vec{I}_h \cdot \vec{S}_{\ell h}\rangle_j$ as 
\begin{equation}
\mbox{
\begin{tabular}{c|c|c}
$i,j$ & 1 & 2 \\[2mm]
\hline
1 & $\displaystyle -\frac{2k_\ell+3}{4(2k_\ell+1)}$ & 
$\displaystyle \frac{\sqrt{k_\ell(k_\ell+1)}}{2k_\ell+1}$ \\[2mm]
\hline
2 & $\displaystyle \frac{\sqrt{k_\ell(k_\ell+1)}}{2k_\ell+1}$ 
  & $\displaystyle - \frac{2k_\ell-1}{4(2k_\ell+1)}$ \\
\end{tabular}
} 
\hskip 5mm
\mbox{
\begin{tabular}{c|c|c}
$i,j$ & 3 & 4 \\[2mm]
\hline
3 & $\frac14$ & 0 \\[2mm]
\hline
4 & 0 & $\frac14$ \\
\end{tabular}
} 
\end{equation}

Subtracting $\frac13(\vec{I}_h \cdot \vec{S}_{\ell h})$, 
the third operator can be expressed in a tensor operator form of
\begin{equation}
{\cal O}_T \equiv (\vec{I}_h \cdot \hat{\vec{x}})
(\vec{S}_{\ell h} \cdot \hat{\vec{x}})
 - \textstyle\frac13(\vec{I}_h \cdot \vec{S}_{\ell h}).
\end{equation}
By rewriting it as 
\begin{equation}
{\cal O}_T = 2 (\vec{I}_h \cdot \hat{\vec{x}}) 
(\vec{I}_h \cdot \vec{S}_{\ell h}) 
(\vec{I}_h \cdot \hat{\vec{x}})
+  \textstyle\frac16(\vec{I}_h \cdot \vec{S}_{\ell h}), 
\end{equation} 
which enables us to use the relations 
$ 2(\vec{I}_h \cdot \hat{x}) |k_\ell, m\rangle_{i} 
= - |k_\ell, m\rangle_{i+2}$ ($i$=1,2), 
we obtain ${}_i \langle {\cal O}_T \rangle_j$ as
\begin{equation}
\mbox{
\begin{tabular}{c|c|c}
$i,j$ & 1 & 2 \\[2mm]
\hline
1 & $\displaystyle \frac{k_\ell}{6(2k_\ell+1)}$ & 
$\displaystyle \frac{\sqrt{k_\ell(k_\ell+1)}}{6(2k_\ell+1)}$ \\[2mm]
\hline
2 & $\displaystyle \frac{\sqrt{k_\ell(k_\ell+1)}}{6(2k_\ell+1)}$ 
  & $\displaystyle \frac{k_\ell+1}{6(2k_\ell+1)}$ \\
\end{tabular}
} 
\hskip 5mm
\mbox{
\begin{tabular}{c|c|c}
$i,j$ & 3 & 4 \\[2mm]
\hline
3 & $\displaystyle -\frac{k_\ell+2}{6(2k_\ell+1)}$ & 
$\displaystyle \frac{\sqrt{k_\ell(k_\ell+1)}}{2(2k_\ell+1)}$ \\[2mm]
\hline
2 & $\displaystyle \frac{\sqrt{k_\ell(k_\ell+1)}}{2(2k_\ell+1)}$ 
  & $\displaystyle - \frac{k_\ell-1}{6(2k_\ell+1)}$ \\
\end{tabular}
} 
\end{equation}

Finally, we obtain $V_{ij}(i,j$=1,2 or 3,4) as 
\begin{equation}
\begin{array}{l}
\displaystyle
V_{11} = V_r + V_{\rm cent} + \frac{k_\ell}{2} V_{IL} -
\frac{2k_\ell +3}{4(2k_\ell+1)} V_{IS} + \frac{k_\ell}{6(2k_\ell+1)}
V_T,  \\[3mm]
\displaystyle
V_{12} (=V_{21}) = \frac{\sqrt{k_\ell(k_\ell+1)}}{2k_\ell+1} V_{IS} +
\frac{\sqrt{k_\ell(k_\ell+1)}}{6(2k_\ell+1)} V_T, \\[3mm]
\displaystyle
V_{22} =  V_r + V_{\rm cent} - \frac{k_\ell+1}{2} V_{IL} -
\frac{2k_\ell-1}{4(2k_\ell+1)} V_{IS} + \frac{k_\ell+1}{6(2k_\ell+1)} V_T,
\\[3mm]
\displaystyle
V_{33} = V_r + V_{\rm cent} - \frac{k_\ell+2}{2} V_{IL} + \frac14 V_{IS}
- \frac{k_\ell+2}{6(2k_\ell+1)} V_T, \\[3mm]
\displaystyle
V_{34} (=V_{43}) = \frac{\sqrt{k_\ell(k_\ell+1)}}{2(2k_\ell+1)} V_T, \\[3mm]
\displaystyle
V_{44} = V_r + V_{\rm cent} + \frac{k_\ell-1}{2} V_{IL} + 
\textstyle \frac14 V_{IS} \displaystyle
- \frac{k_\ell-1}{6(2k_\ell+1)} V_T.
\end{array}
\end{equation}
where $V_r = \upsilon^2/M_s$, 
$V_{\rm cent} = \ell(\ell+1)/(2M_s x^2)$,
$V_{IL} = 2\upsilon/(M_s x)$, 
$V_{IS} = 2 g_Q ( a_1 + \frac13 a_2 )$
and $V_T = 2g_Q a_2$.

\setcounter{equation}{0}
\section{Numerical Process}

Here, we briefly describe the numerical process of finding the 
bound states.
Equation (\ref{xeq}) can be solved like any other eigenvalue problem. 
First, we find the asymptotic solutions near the origin and at 
large $x$. Except the $k_\ell=0$ case where only one radial function 
($\psi_1(x)$ or $\psi_3(x)$) is defined, 
we can find two independent solution sets. (See Table \ref{t2}.)
Taking those independent sets of asymptotic solutions as initial 
conditions, the differential equation can be solved numerically 
in the forward or backward direction of $x$. 
In this way, we can obtain 8 radial functions (2 independent 
radial functions $\times$ 2 independent initial conditions 
$\times$ integrated forward or backward direction of $x$). 
Let us distinguish those radial functions by using scripts 
$>$(or $<$) and $A$(or $B$). 
For example, $\psi^{>}_{1,A}$ denotes the radial function 
obtained by starting from the asymptotic solution set $A$ 
and solving the equation of motion in backward direction.

Then, a general solution that one can obtain by integrating 
the equation of motion in forward direction is expressed as 
\begin{equation}
\left( \begin{array}{c} 
\psi_1^{<}(x) \\ \psi_2^{<}(x) 
\end{array} \right) = 
\left( \begin{array}{c}
\alpha_A^{} \psi_{1,A}^{<} + \alpha_B^{} \psi_{1,B}^{<} \\
\alpha_A^{} \psi_{2,A}^{<} + \alpha_B^{} \psi_{2,B}^{<} 
\end{array} \right) , 
\end{equation}
where $\alpha_{A,B}^{}$ are arbitrary constants. 
Similarly, a general backward solution can be obtained as
\begin{equation}
\left( \begin{array}{c} 
\psi_1^{>}(x) \\ \psi_2^{>}(x) 
\end{array} \right) = 
\left( \begin{array}{c}
\beta_A^{} \psi_{1,A}^{>} + \beta_B^{} \psi_{1,B}^{>} \\
\beta_A^{} \psi_{2,A}^{>} + \beta_B^{} \psi_{2,B}^{>} 
\end{array} \right) .
\end{equation}

\begin{table} 
\caption{\sl Two independent asymptotic solutions near the origin 
and at large $x$.}
\vskip 3mm \begin{center}
\begin{tabular}{cllll}
\hline\hline
 & \multicolumn{2}{c}{$\pi=-(-1)^{k_\ell}$} 
 & \multicolumn{2}{c}{$\pi=+(-1)^{k_\ell}$} \\
 & \multicolumn{1}{c}{near the origin} & \multicolumn{1}{c}{at large $x$} 
 & \multicolumn{1}{c}{near the origin} & \multicolumn{1}{c}{at large $x$} \\
\hline
Sol. A  
 & $\begin{array}{l} \psi_1(x) \sim x^{k_\ell+1} \\  
    \psi_2(x) \sim 0 \end{array}$ 
 & $\begin{array}{l} \psi_1(x) \sim e^{\sqrt{|\varepsilon_n|} x}/x \\
    \psi_2(x) \sim 0 \end{array}$ 
 & $\begin{array}{l} \psi_3(x) \sim x^{k_\ell} \\  
    \psi_4(x) \sim 0 \end{array}$ 
 & $\begin{array}{l} \psi_3(x) \sim e^{\sqrt{|\varepsilon_n|} x}/x \\
    \psi_4(x) \sim 0 \end{array}$ \\
\hline
Sol. B 
 & $\begin{array}{l} \psi_1(x) \sim 0 \\
    \psi_2(x) \sim x^{k_\ell-1} \end{array}$ 
 & $\begin{array}{l} \psi_1(x) \sim 0  \\
    \psi_2(x) \sim e^{\sqrt{|\varepsilon_n|} x}/x \end{array}$ 
 & $\begin{array}{l} \psi_3(x) \sim 0 \\
    \psi_4(x) \sim x^{k_\ell} \end{array}$ 
 & $\begin{array}{l} \psi_3(x) \sim 0  \\
    \psi_4(x) \sim e^{\sqrt{|\varepsilon_n|} x}/x \end{array}$ \\
\hline\hline
\end{tabular}
\end{center}\label{t2}
\end{table}   

At a proper intermediate position $x_m$, 
those wavefunctions should be matched in the way 
that the functions and their derivatives are continuous. 
This matching condition leads to a linear equation,
\begin{equation}
\left( \begin{array}{cccc} 
\psi^{<}_{1,A}(x_m) & \psi^{<}_{1,B}(x_m) & 
  \psi^{>}_{1,A}(x_m) & \psi^{>}_{1,B}(x_m) \\ 
\psi^{<}_{2,A}(x_m) & \psi^{<}_{2,B}(x_m) & 
  \psi^{>}_{2,A}(x_m) & \psi^{>}_{2,B}(x_m) \\ 
\psi^{<\prime}_{1,A}(x_m) & \psi^{<\prime}_{1,B}(x_m) & 
  \psi^{>\prime}_{1,A}(x_m) & \psi^{>\prime}_{1,B}(x_m) \\ 
\psi^{<\prime}_{2,A}(x_m) & \psi^{<\prime}_{2,B}(x_m) & 
  \psi^{>\prime}_{2,A}(x_m) & \psi^{>\prime}_{2,B}(x_m) 
\end{array}\right) 
\left( \begin{array}{r} 
\alpha_A \\ \alpha_B \\ -\beta_A \\ -\beta_B 
\end{array} \right) = 0.
\label{leq}\end{equation}
In order that it supports a nontrivial solution, the determinant of the 
matrix should vanish. Such a secular equation determines the 
eigenenergies. Once an eigenenergy is found, Eq. (\ref{leq}) 
can be solved for the ratios between the coefficients, 
$\alpha_{A,B}$ and $\beta_{A,B}$. 
Their absolute values can be fixed by normalizing the wavefunction.

In the special case of $k_\ell=0$, there appears only one radial 
function, $\psi_1(x)$ or $\psi_3(x)$. The asymptotic solution 
of them is given as the solution set A in Table \ref{t3}. 
In a similar way described above, one obtains 
the backward and forward solutions, $\psi^{>}_{1(3)}$ and 
$\psi^{<}_{1(3)}$, respectively. 
Simply by matching their logarithmic derivatives as
\begin{equation}
\left. \frac{\psi^{<\prime}_{1(3)}}{\psi^{<}_{1(3)}} \right|_{x=x_m} = 
\left. \frac{\psi^{>\prime}_{1(3)}}{\psi^{>}_{1(3)}} \right|_{x=x_m} ,
\end{equation}
one can find the eigenstates.


\end{document}